\documentclass[draft]{agujournal2019}
\usepackage{url} 
\usepackage{lineno}
\usepackage{amsmath}
\usepackage[inline]{trackchanges} 
\usepackage{soul}
\draftfalse

\journalname{JGR: Space Physics}

\begin{document}

%
%


\title{Differentiating EDRs from the Background Magnetopause Current Sheet: A Statistical Study}

%
%




\authors{J. M. H. Beedle\affil{1,2}, D. J. Gershman\affil{2}, V. M. Uritsky\affil{1,2}, T. D. Phan\affil{3}, B. L. Giles\affil{2}}

\affiliation{1}{Department of Physics, The Catholic University of America, Washington D.C. USA}
\affiliation{2}{NASA Goddard Space Flight Center, Greenbelt, MD, USA}
\affiliation{3}{Space Sciences Laboratory, University of California, Berkeley, CA, USA}




\correspondingauthor{Jason M. H. Beedle}{beedle@cua.edu}




\begin{keypoints}
\item EDR crossings show current densities an order of magnitude higher than regular magnetopause crossings.
\item EDR crossings contain significant current components parallel to the local magnetic field. 
\item Ion velocity along the magnetopause is highly correlated with an event's location - indicating the presence of magnetosheath flows.
\end{keypoints}

%
%

%
%

\begin{abstract}

The solar wind is a continuous outflow of charged particles from the Sun's atmosphere into the solar system. At Earth, the solar wind's outward pressure is balanced by the Earth's magnetic field in a boundary layer known as the magnetopause. Plasma density and temperature differences across the boundary layer generate the Chapman-Ferraro current which supports the magnetopause. Along the dayside magnetopause, magnetic reconnection can occur in electron diffusion regions (EDRs) embedded into the larger ion diffusion regions (IDRs). These diffusion regions form when opposing magnetic field lines in the solar wind and Earth’s magnetic field merge, releasing magnetic energy into the surrounding plasma. While previous studies have given us a general understanding of the structure of the diffusion regions, we still do not have a good grasp of how they are statistically differentiated from the non-diffusion region magnetopause. By investigating 251 magnetopause crossings from NASA’s Magnetospheric Multiscale (MMS) Mission, we demonstrate that EDR magnetopause crossings show current densities an order of magnitude higher than regular magnetopause crossings - crossings that either passed through the reconnection exhausts or through the non-reconnecting magnetopause, providing a baseline for the magnetopause current sheet under a wide range of driving conditions. Significant current signatures parallel to the local magnetic field in EDR crossings are also identified, which is in contrast to the dominantly perpendicular current found in the regular magnetopause. Additionally, we show that the ion velocity along the magnetopause is highly correlated with a crossing's location, indicating the presence of magnetosheath flows inside the magnetopause.

\end{abstract}

\section*{Plain Language Summary}

The magnetopause is a dynamic boundary layer created through the interaction of the solar wind with Earth's magnetic field. This boundary is supported by a current sheet and acts as the ``entry gate" of the solar wind's energy into the magnetosphere through a process called magnetic reconnection where energy previously stored in the magnetic field is released into the surrounding magnetopause plasma. The reconnection process is initiated in localized diffusion regions, which form in the magnetopause's current sheet during specific solar wind conditions. In this paper, we clarify what makes the diffusion regions stand out from the background magnetopause current sheet by utilizing data from NASA's Magnetospheric Multiscale mission. Our analysis reveals that the diffusion regions have stronger currents than the background magnetopause and that a significant portion of this current becomes parallel to the local magnetic field.

%
%

%


%
%
%
%

\section{Introduction}

The magnetopause is a boundary layer created through the balancing of the solar wind’s dynamic pressure with Earth’s magnetic field. Across this boundary layer, pressure gradients generate a current sheet, named the Chapman-Ferraro (CF) current, that supports the magnetopause (e.g. \citeA{CF1931}). This current sheet is a large scale, mainly {ion-driven} current generated from ion density and temperature gradients - {e.g. \citeA{Hasegawa2012} and references therein}. 

Along the dayside magnetopause current sheet, magnetic reconnection occurs when opposing field lines in the solar wind and Earth’s magnetic field are driven together by plasma flows, causing the magnetopause boundary to thin {- e.g. \citeA{TreumannBaumjohann2013}}. As the current sheet compresses, there becomes a small-scale region where the frozen-in condition in the plasma is violated, allowing the magnetic field to become disassociated from the plasma and diffuse through it, break, and then reform, changing the local magnetic topology (e.g. \citeA{Vasyliunas1975, HesseCassak2020}). This process occurs in what are known as diffusion regions. Specifically, there are two distinct regions: an ion diffusion region (IDR) and an electron diffusion region (EDR). The IDR is the larger outer region where ions first dissociate from the magnetic field while the electrons remain frozen-in. This hybrid configuration with magnetized electrons and free ions then creates the Hall currents and their associated quadrupole Hall magnetic field in the IDR (e.g. \citeA{Sonnerup1979,Oieroset2001,Mozer2002}). The EDR is the smaller inner diffusion region, embedded in the larger IDR, where both the electrons and ions are decoupled from the magnetic field, which allows magnetic reconnection to take place - e.g. \citeA{Vasyliunas1975}, \citeA{Burch2016}. 

The process of magnetic reconnection leads to the magnetopause acting as the ``entry gate” of  the solar wind's energy into the Earth’s magnetosphere. Thus a thorough grasp of this process and its impact on the magnetopause’s current sheet is vital to understanding the energy transfer into the terrestrial space weather system. Because of this significance, numerous missions {(International Sun-Earth Explorer - ISEE, Active Magnetospheric Particle Tracer Explorers - AMPTE, Polar, Cluster, Time History of Events and Macroscale Interactions during Substorms - Themis, Magnetospheric Multiscale Mission - MMS, etc)} devoted their resources to gaining insights into the magnetopause current sheet and dayside magnetopause reconnection. {A number of studies [\citeA{Burch2016}, \citeA{Lavraud2016}, \citeA{Norgren2016}, \citeA{Phan2016}, \citeA{BurchPhan2016}, \citeA{Chen2016}, \citeA{Chen2017}, etc.]} have focused on individual dayside magnetopause EDR events using the MMS spacecraft to study common elements of EDR crossings including ion jets and jet reversals around the diffusion regions, plasma inflows, non-gyrotropic crescent shaped electron outflows, intense currents, and strong heating. {Other studies [\citeA{Rager2018}, \citeA{Webster2018}, \citeA{Shuster2019}, \citeA{Genestreti20}, \citeA{Shuster2021}, etc.]} have addressed these generalized characteristics of EDR events in more detail and confirmed the prevalence of crescent-shaped electron velocity phase space densities, ohmic heating of the plasma, as well as the role of electron scale currents to the {EDR}. 

{While previous studies provide a general understanding of the diffusion regions' structure, there has not yet been a statistical study of the characteristic differences between magnetopause crossings with and without active signatures of reconnection.} To begin answering this question, our study analyzes data from NASA’s MMS Mission during EDR and {regular} magnetopause crossings, where the {regular} events could either be encounters of the reconnection exhausts downstream of the diffusion region, or non-reconnecting magnetopause crossings{, providing a baseline for the magnetopause current sheet under a wide range of solar wind driving conditions}. In Section 2, we describe the methods we used to accomplish this analysis as well as the findings from comparing the EDR and {regular} magnetopause crossings. Section 3 covers an in-depth analysis of our results, with observations about the magnetopause's current structure and ion velocities measured during these EDR events. Section 4 introduces a brief discussion of our findings, and Section 5 provides a summary of the main results of this study. 

\section{Observations}

\subsection{MMS Data and Current Calculations}

For this study, we utilized data from NASA's MMS mission, which is a mission comprised of four spacecraft that travel in a tetrahedron pattern through the magnetosphere. MMS’s Fast Plasma Investigation (FPI) \cite{FPI} and Fluxgate Magnetometer \cite{FGM} instruments enable four simultaneous measurements of plasma properties and magnetic field conditions, respectively, across MMS’s constellation. Using the data from these two instruments, we analyzed the magnetopause current system through the following currents. 

The first is called the curlometer current, or $\textbf{J}_{curl}$, which was calculated using \citeA{Dunlop1988}'s curlometer method to approximate gradients in the magnetic field, yielding Ampere’s law in the MHD approximation:

\begin{equation}
	\textbf{J}_{curl} = \frac{\nabla \times \textbf{B}}{\mu_0},
\end{equation}

\noindent {where $\textbf{B}$ is the magnetic field and $\mu_o$ is the permeability of free space.} $\textbf{J}_{curl}$ represents the current consistent with magnetic field perturbations and is thus a proxy for the total current density encountered by MMS during a magnetopause current sheet crossing as it contains current components parallel and perpendicular to the magnetic field including both ion and electron contributions. While $\textbf{J}_{curl}$ has been found to be less sensitive to small current structures than the current density calculated from plasma moments, also known as the FPI current, we decided to focus on using $\textbf{J}_{curl}$ as the average MMS separation during our time frame (2015–2018) was between {10 to 60 kms}, which is sufficiently close to consider the {ion-dominated} CF current as well as any large scale currents in and around the diffusion regions in EDR events. This is in contrast to studies that have focused more heavily on the smaller scale, {electron-dominated} currents of the EDR, which generally use the FPI current (e.g. \citeA{Lavraud2016}; \citeA{Phan2016}). {Additionally, the curlometer current allows for a better comparison with diamagnetic currents which are estimated by using multi-spacecraft gradient measurements, thus meaning that both currents measure over the tetrahedron of the MMS spacecraft and under-resolve substructures in the same manner.} We also considered the components of $\textbf{J}_{curl}$ parallel and perpendicular to the locally measured magnetic field $\textbf{B}$: $\textbf{J}_{curl \parallel}$ and $\textbf{J}_{curl \perp}$, defined as follows:

\begin{equation}
	{\textbf{J}_{curl \parallel} = \left( \frac{\textbf{B} \cdot \textbf{J}_{curl}}{|\textbf{B}|} \right) \hat{\textbf{B}} \ \ , \ \  
	\textbf{J}_{curl \perp} = \textbf{J}_{curl} - \textbf{J}_{curl \parallel}}.
\end{equation}

\noindent Note, $\textbf{B}$ is averaged across all four MMS spacecraft to match the curlometer method calculations.

Along with $\textbf{J}_{curl}$, we looked at the ion and electron diamagnetic currents: $\textbf{J}_{dia \ Total_i}$ and $\textbf{J}_{dia \ Total_e}$ and their current components generated from temperature and density gradients, which were approximated using the curlometer method. Both the ion and electron diamagnetic currents and their density and temperature components were defined in the following manner, with their respective densities and temperatures, in the same way as \citeA{Beedle2022}:

\begin{equation}
	\textbf{J}_{dia \  \nabla N} = \frac{\textbf{B}\times(k_b\overleftrightarrow{T} \cdot \nabla N)}{|\textbf{B}|^2} \ \ , \ \  
	\textbf{J}_{dia \ \nabla \cdot \overleftrightarrow{T}} = \frac{\textbf{B}\times(k_bN\nabla \cdot \overleftrightarrow{T})}{|\textbf{B}|^2},
\end{equation}

\noindent {where $\textbf{B}$ represents the magnetic field, $k_b$ is Boltzmann's constant, $\overleftrightarrow{T}$ is the temperature tensor, and N is the number density.} By definition, $\textbf{J}_{dia \ Total} = \textbf{J}_{ dia \  \nabla N} + \textbf{J}_{dia \ \nabla \cdot \overleftrightarrow{T}}$. Note, $\textbf{B}$, $\overleftrightarrow{T}$, and N were averaged across all four MMS spacecraft to match with the use of the curlometer method to calculate the gradients. When referencing these components, we will refer to $\textbf{J}_{ dia \  \nabla N}$ as its current's density component and to $\textbf{J}_{dia \ \nabla \cdot \overleftrightarrow{T}}$ as its current's temperature component.

While we considered both parallel and perpendicular components for $\textbf{J}_{curl}$, diamagnetic current is, by definition, perpendicular to the magnetic field, thus the ion and electron diamagnetic currents represent the primary perpendicular components to the magnetic field in the magnetopause.

To summarize, we analyzed the following set of current densities: $\textbf{J}_{curl}$, $\textbf{J}_{dia \ Total_i}$, $\textbf{J}_{dia \ Total_e}$, $\textbf{J}_{dia \ \nabla N_i}$, $\textbf{J}_{dia \ \nabla \cdot T_i}$, $\textbf{J}_{dia \ \nabla N_e}$, $\textbf{J}_{dia \ \nabla \cdot T_e}$ during each of the studied magnetopause crossings.

When interpreting these quantities, it is important to note that MHD physics breaks down in the diffusion regions as plasma disassociates from the magnetic field. Because the diamagnetic current equations are defined under MHD conditions, the concept of a diamagnetic current also breaks in the diffusion regions as the plasma must now be described using kinetic theories. Thus, while the current around the diffusion region is still represented by the diamagnetic, ion dominated CF current, inside the diffusion regions they become kinetic and can no longer be described in the same way. For this reason our results using the diamgnetic current are more likely to contain {anomalously large current spikes} once the MMS constellation entered the IDR and EDR of that magnetopause crossing event. However, as the EDR itself is still quite small when compared with the current sheet that MMS observes, there are regions where the diamgnetic current is a useful measure. {Specifically, the sizes of these regions are estimated to be on the order of 50 km for the IDR and 1 km for the EDR - e.g. \citeA{Phan2015}. Studies such as \citeA{Torbert2017} have reinforced these values by identifying an EDR with a size of approximately 2 km, which appeared as roughly 40 ms of MMS data. From the EDR events that we analyzed in this study, the magnetopause crossings lasted from a low of 1.8 seconds to a high of almost 17 seconds. In either case, the magnetopause boundary captured covers a significantly larger time span than the brief EDR, and is at least twice as large than the IDR in the limiting case. Of course, how the MMS constellation cuts across each specific event significantly complicates this process, but a more in-depth timing analysis is out of the scope of this study.} It is also worthy to note that, as the curlometer current relies on the deviations in the magnetic field itself, it is not impacted in the same manner and presents accurate current measurements all throughout the current sheet crossing, be it in the current sheet itself, or the diffusion regions.

{Data taken from MMS, as well as the calculated currents, were interpolated to the 30 ms FPI electron time resolution from the 150 ms ion time resolution and the 10 ms magnetometer time resolution following standard practice in MMS studies - see \citeA{Burch2016}, \citeA{Phan2016}, etc.} As our main analysis involves ion and electron diamagnetic currents and the total current as computed from the curlometer method, any sub 150 ms variations in the ion parameters should not impact our results. For all measured quantities that did not use the curlometer method, we averaged over all four MMS spacecraft to create a single data stream. Our calculations and measurements were completed in Cartesian GSE coordinates and then stored in spherical GSE coordinates, in which the $\phi$ angle is in the primary current direction along the dayside magnetopause as can be seen defined in Figure \ref{fig:Event_locations}. A more detailed description of these spherical coordinates is provided in Section 3.

\subsection{Event Selection}

\begin{figure}[htbp]
    \centering
    \noindent\includegraphics[width=0.9\textwidth]{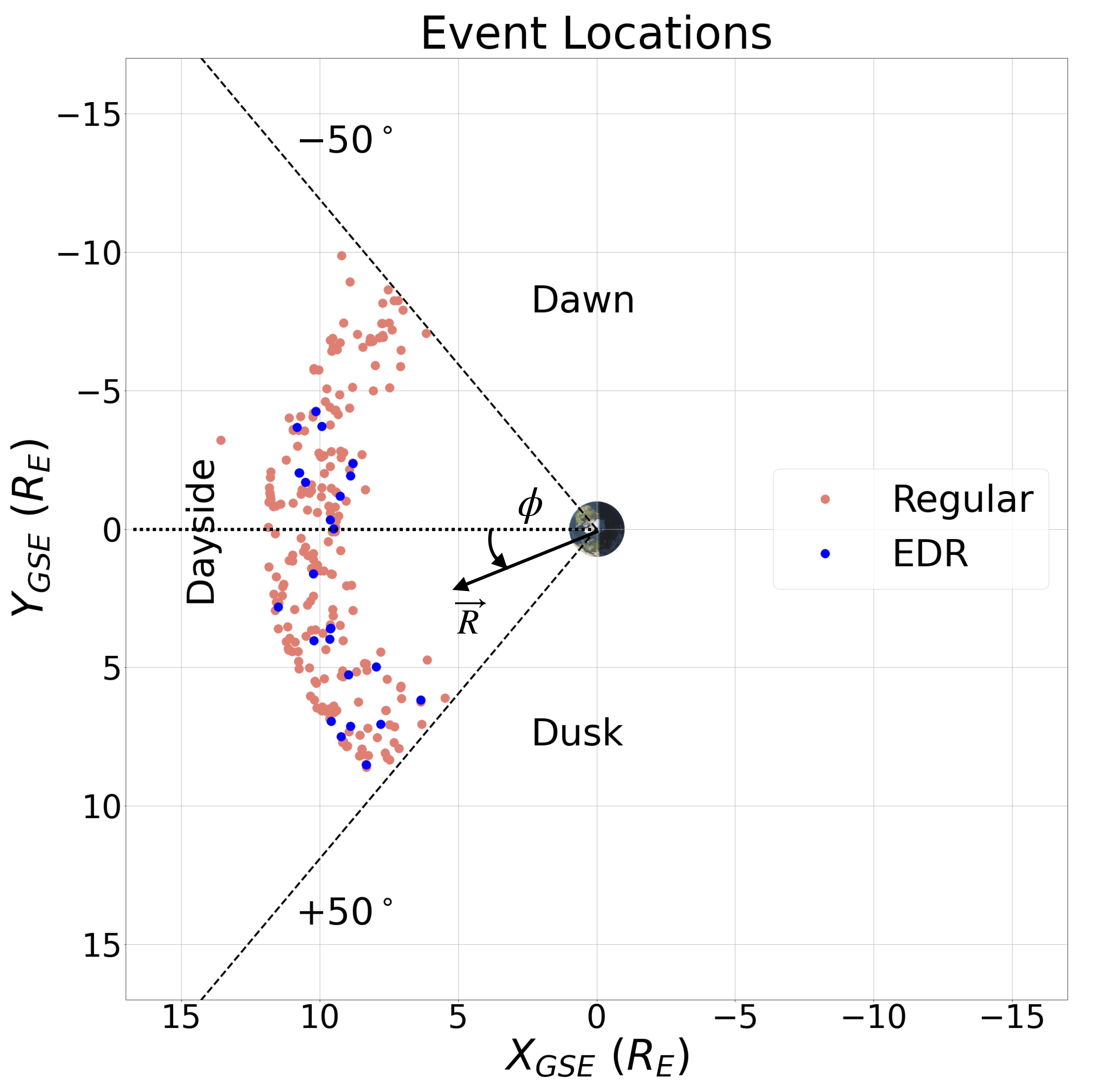}
    \caption{Diagram of our 225 dayside {regular} (red) and 26 EDR (blue) magnetopause crossings. We define a spherical coordinate system with $\phi$ in the $X_{GSE}$ - $Y_{GSE}$ plane, positively defined from the $+X_{GSE}$ axis, R defined as radially outward,  and $\theta$ as the {polar} angle into the $+Z_{GSE}$ direction, completing the right-handed coordinate system. The Dayside is defined as being from $+50^{\circ}$ to $-50^{\circ}$ in $\phi$, following the same convention as \citeA{Beedle2022}.}
    \label{fig:Event_locations}
\end{figure}

{To select relevant data for our study, we used EDR crossings provided by \citeA{Webster2018}, who compiled previously identified EDR events with a set of newly-identified EDR events based on shared characteristics including the occurrence of non-gyrotropic crescent-shape electron distributions, ion jet reversals, and large current densities. Because of this reliance on non-gyrotropic electron distributions to identify EDR events, the \citeA{Webster2018} events can only include, at most, a moderate guide field as stronger guide fields tend to obscure this feature {\cite{Hesse2011,Genestreti2017}}. In all, \citeA{Webster2018} reported 32 EDR events, 26 of which were included in our study based on their location along the dayside magnetopause as well as the availability of MMS data from all four spacecraft. Four of Webster et al.'s events (A13, B14, B15, and B17) were located outside of the bounds of our definition of the dayside magnetopause (see Figure \ref{fig:Event_locations}), while two other events (A7 and B26) caused errors with our code because of data outages from one or more MMS spacecraft. The selected 26 EDR events then represented the EDR sample group that we measured the aforementioned current densities and other plasma characteristics over. The locations of these events along the dayside magnetopause are denoted in blue in Figure \ref{fig:Event_locations}.} 

\begin{figure}[htbp]
    \centering
    \noindent\includegraphics[width=1.0\textwidth]{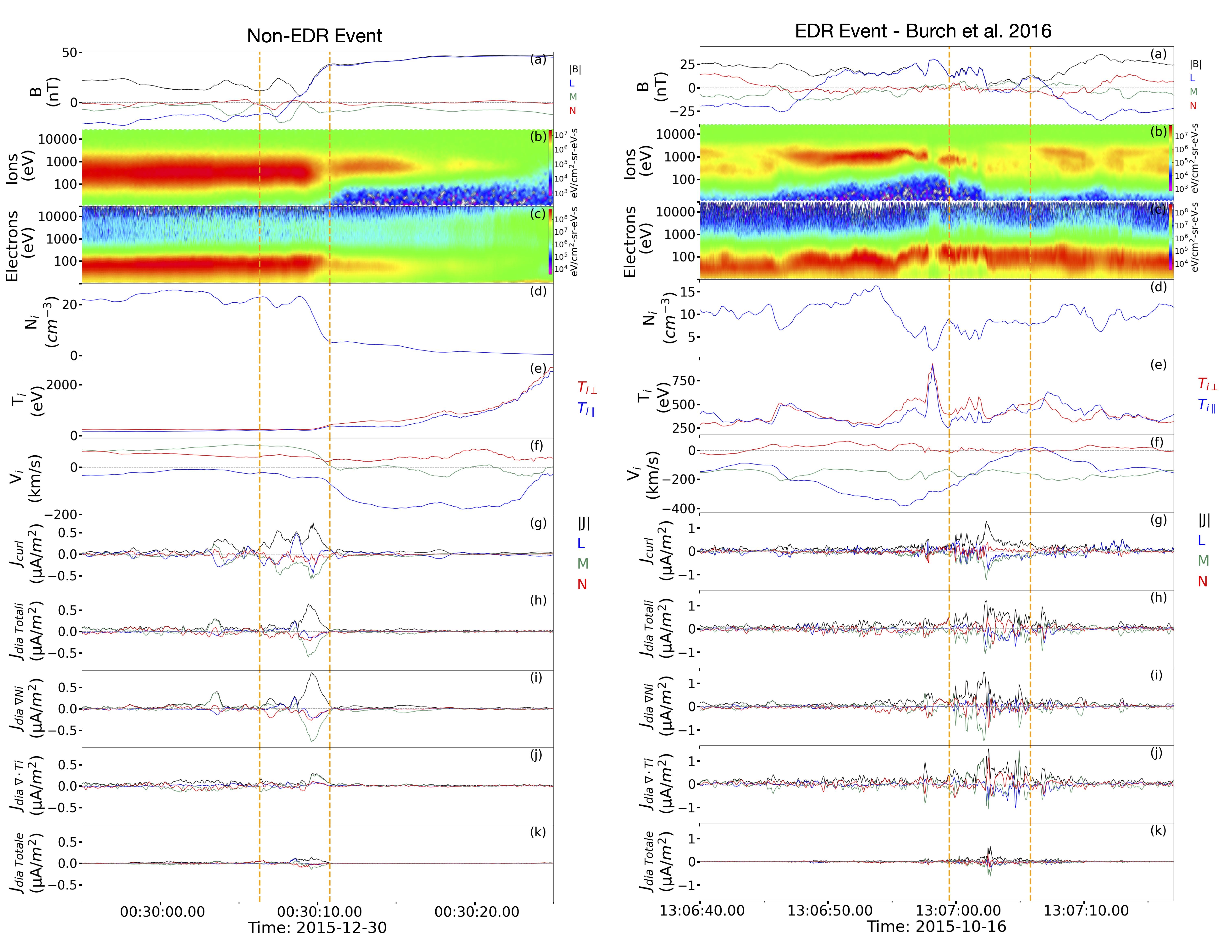}
    \caption{Example magnetopause crossings representing a {regular} event (left) and an EDR event (right). The orange dashed lines represent the magnetopause crossing as identified by our algorithm for each event (see Section 2.3). The example EDR event is from Burch et al. 2016. (a) Magnitude and magnetic field in LMN coordinates {determined through MVAB analysis \cite{Sonnerup1998}}, (b) and (c) ion and electron omni directional spectrograms, (d) ion number density, (e) ion perpendicular and parallel temperature, (f) ion velocity, (g, h, i, j and k) curlometer, total ion diamagnetic, ion density component, ion temperature component, and total electron diamagnetic current densities respectively in LMN coordinates with magnitudes indicated in black.}
    \label{fig:Event_Figure}
\end{figure}

Along with these 26 EDR events, we also investigated 225 dayside magnetopause current sheet crossings taken from \citeA{Paschmann2018} and \citeA{Haaland2020}’s database of MMS magnetopause crossings. {These 225 events were previously used in the \citeA{Beedle2022} study and include complete, monotomic magnetopause crossings. In \citeA{Beedle2022}, a small number of events with unusually high (above $2,000 nA/m^2$) current density were manually removed in order to reduce the possibility of including a reconnection event as the MMS database itself includes all manner of magnetopause crossing and may include a small subset of previously unrecognized, and unpublished EDR crossings. Note that none of our 26 EDR events are included in the 225 events from the database.} These 225 events then represent our {regular} magnetopause crossing sample group that we compared with the EDR samples. As previously mentioned, these {regular} crossings could either be a crossing of the reconnection exhausts downstream of the diffusion regions, or a crossing of the non-reconnection magnetopause{, providing a baseline for the magnetopause current sheet under a range of solar wind driving conditions.} \citeA{Beedle2022} provides a detailed explanation of the selection criteria for these 225 events. The locations of the 225 magnetopause crossing events are denoted in red in Figure \ref{fig:Event_locations}. An example of a {regular} magnetopause crossing versus an EDR crossing is provided in Figure \ref{fig:Event_Figure}. 

\subsection{Magnetopause Identification}

Each of the events in our database was processed by an algorithm to identify their magnetopause crossing times. We used $\textbf{J}_{curl}$ to identify the largest current magnitude peak during an event, and then applied a threshold equal to $20\%$ of this peak value to the current density measured during the crossing. This separated the event into current segments, with each weighted based on their duration, average $|\textbf{J}_{curl}|$ current density, and the magnetic field magnitude, $|\textbf{B}|$, measured over the segment. The segment with the longest duration, highest average current density, and largest change in $|\textbf{B}|$ across the segment was then selected as the primary current sheet crossing for that event, with the start and end times of the current segment then becoming the beginning and end of that event’s magnetopause crossing. Two examples of the algorithm's selection method can be seen represented by the vertical orange dashed lines in Figure \ref{fig:Event_Figure}. Note, as this method uses the magnitudes of each value, it is coordinate system invariant. 

This method was applied to both the EDR and {regular} crossings, with the results for the average current density for the 225 {regular} events matching within error the average current over the magnetopause crossing times identified by the \citeA{Paschmann2018} database’s minimum variance analysis method as previously reported in \citeA{Beedle2022}. The performance of our algorithm was also manually checked over the 26 EDR events so that the selected magnetopause crossing correctly captured the EDR event as previously identified by their respective papers.

\section{EDR and {Regular} Magnetopause Crossing Analysis}

Over each of the 26 EDR events and 225 {regular} magnetopause crossings, we recorded individual current density data and stored the results in spherical GSE coordinates with R being defined as radially outward, $\phi$ going from dawn-to-dusk in the $X_{GSE}$-$Y_{GSE}$ plane, and $\theta$ pointing in the $+Z_{GSE}$ direction (see Figure \ref{fig:Event_locations} for a visual depiction). {We utilize spherical GSE coordinates instead of LMN coordinates for our statistical survey to be able to compare current density components measured over the EDR and {regular} events on an equal footing. In previous statistical studies (e.g. \citeA{Paschmann2018}, \citeA{Haaland2020}), MVAB analyses \cite{Sonnerup1998} were utilized over MMS's burst mode intervals to generate LMN coordinates for their events. This works well for intervals that involve a single, clear magnetopause crossing, but leads to uncertainties when MMS passes over the magnetopause multiple times in quick succession, such as during active solar wind conditions. These crossings are often nonuniform and contain small scale embedded structure. In such cases, the MVAB analysis interval needs to be adjusted in order to capture the appropriate crossing, which leads to some ambiguity, especially when trying to compare the individual current directions measured over many such events. In the aforementioned studies, the magnitude of the current density was reported for each event, which is unaffected by these differences in coordinate determination. For our statistical survey, as we directly compare currents along coordinate directions, we decided to use a global coordinate system that is equally applied to all of our crossings, regardless of the dynamics involved.} 

Our analysis resulted in 6,332 data points for each current component from the EDR events and 73,865 data points from the {regular} events. We then analyzed the combined data’s mean and median values as well as their standard errors or $\sigma / \sqrt{N}$ where $\sigma$ is the standard deviation of the data and N is the number of data points recorded. {It should be noted that the standard error can be used as a rigorous uncertainty measure only if the values represented by the mean are statistically independent and sampled from a stationary normal distribution. Due to the inherent inhomogeneity of the traversed current density structures, it is possible that these probabilistic assumptions are at least partly violated for some of the crossing events. In view of this, the reported standard deviations should be considered as empirical proxies for the statistical dispersion of the sample means rather than their rigorous uncertainties. The latter could be somewhat smaller or greater than the reported errors depending on the length of the sample and the geometry of the observed structure.}

\begin{figure}[htbp]
    \centering
    \noindent\includegraphics[width=1.0\textwidth]{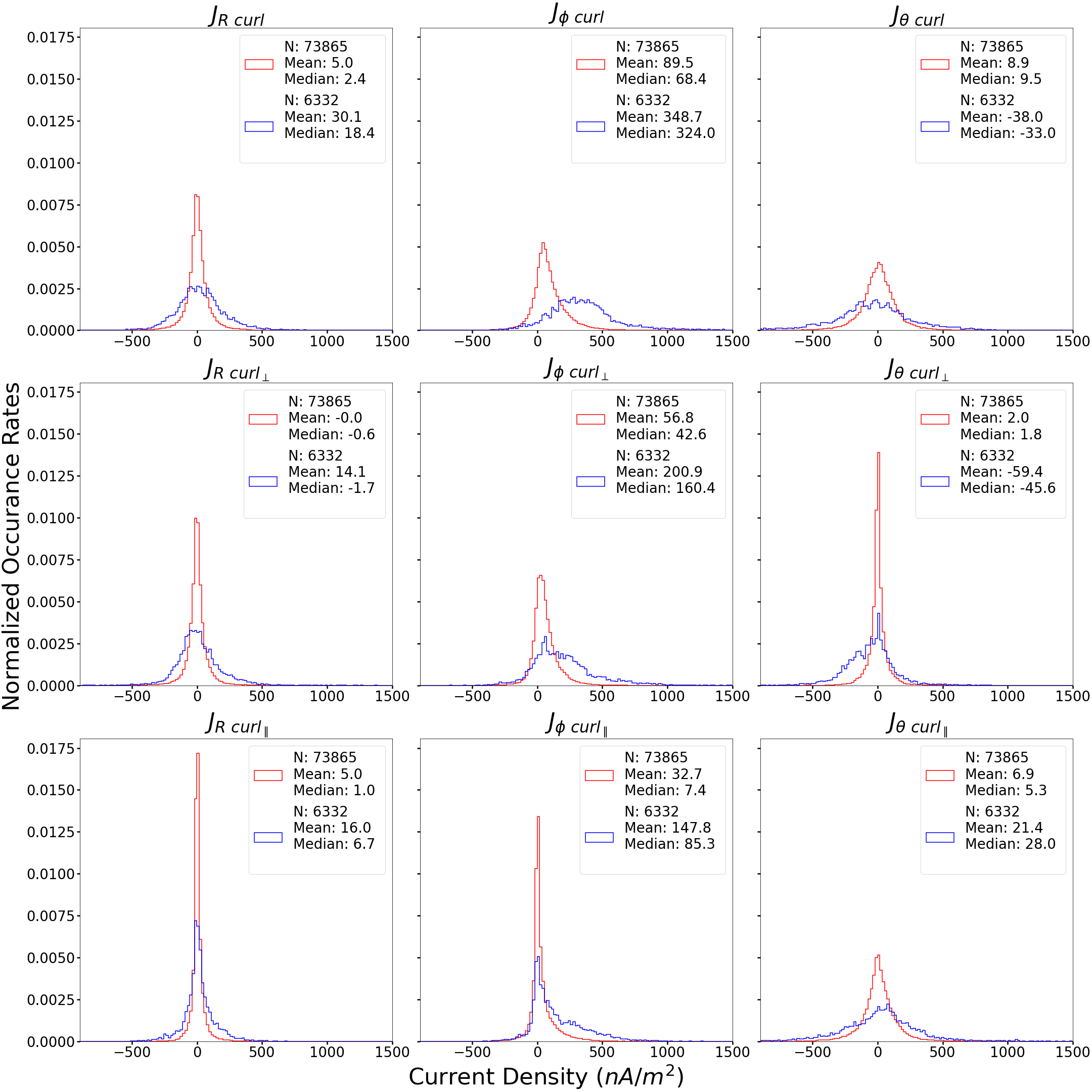}
    \caption{Probability distribution histograms of the curlometer current and its parallel and perpendicular components to the local magnetic field from the 26 EDR crossings (blue) and 225 {regular} magnetopause crossings (red) measured across the three global coordinates ($\hat{R}$, $\hat{\phi}$, $\hat{\theta}$). The EDR events gave us 6,332 data points in total, while the {regular} crossings gave us 73,865 data points. Note that the vertical axis in each plot is normalized, with the same scale used for each subplot for the vertical and horizontal axes respectively. The bins used are also the same for each subplot’s distributions. The sample mean and median values are provided in the top right of each subplot.}
    \label{fig:Jcurl_Hist}
\end{figure}

\begin{figure}[htbp]
    \centering
    \noindent\includegraphics[width=1.0\textwidth]{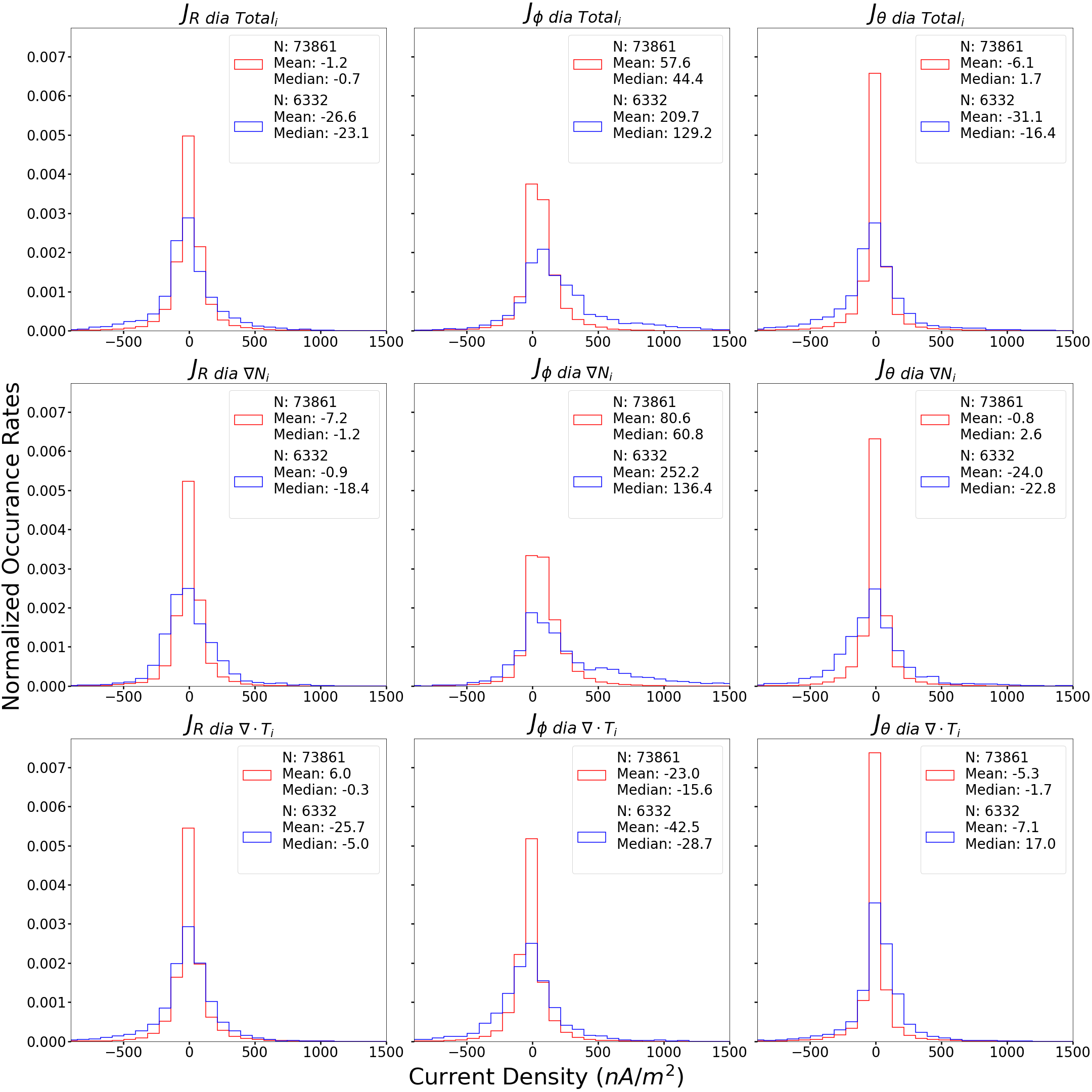}
    \caption{Probability distribution histograms of the ion diamagnetic current and its current components - the density component and the temperature component - in EDR crossings (blue) and {regular} crossings (red) over the spherical $\hat{R}$, $\hat{\phi}$, and $\hat{\theta}$ component directions. The sample mean and median values are provided in the top right of each subplot.}
    \label{fig:ionHist}
\end{figure}

\begin{figure}[htbp]
    \centering
    \noindent\includegraphics[width=1.0\textwidth]{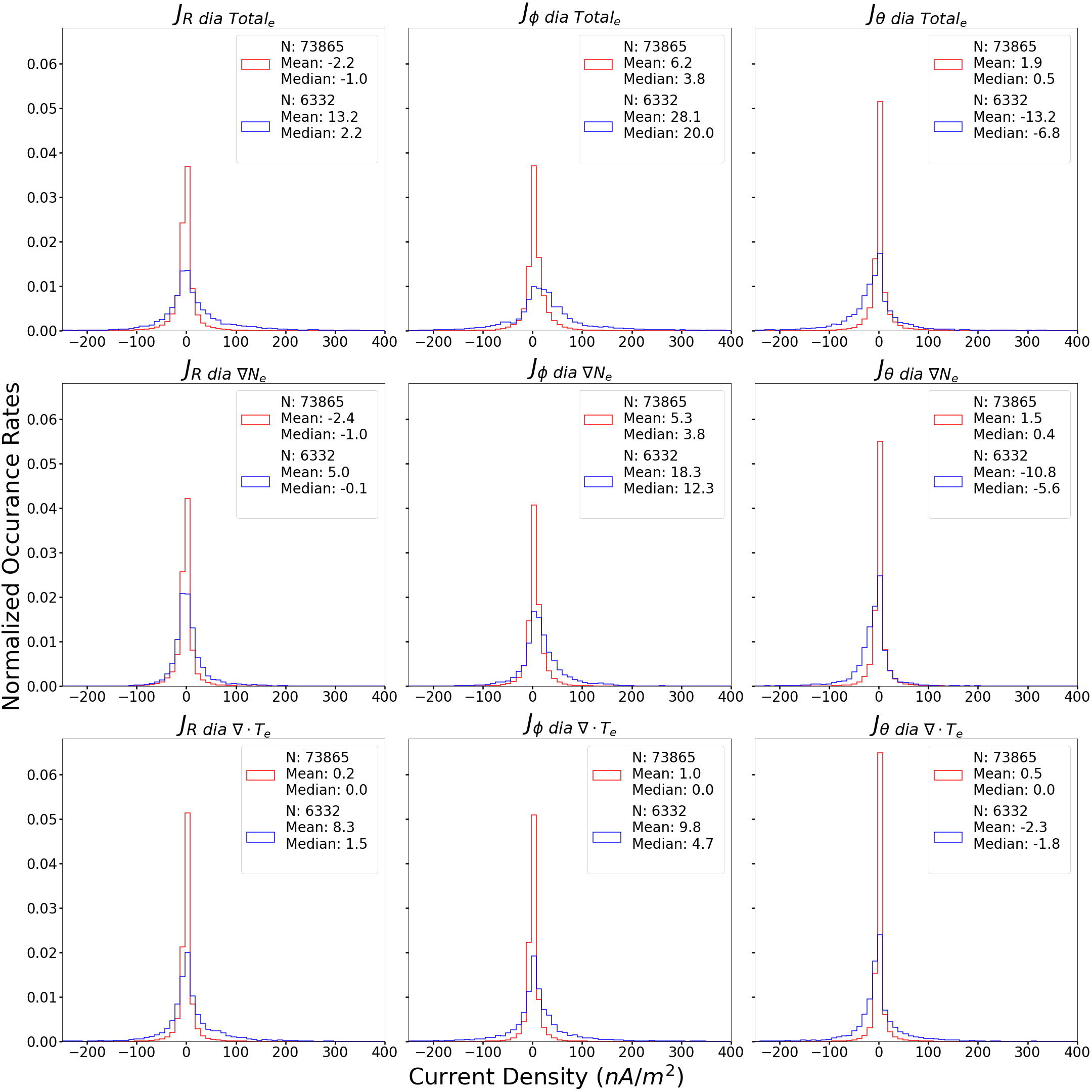}
    \caption{Probability distribution histograms of the electron diamagnetic current and its current components - the density component and the temperature component - in EDR crossings (blue) and {regular} crossings (red) over the spherical $\hat{R}$, $\hat{\phi}$, and $\hat{\theta}$ component directions. The sample mean and median values are provided in the top right of each subplot.}
    \label{fig:elecHist}
\end{figure}

\begin{figure}[htbp]
    \centering
    \noindent\includegraphics[width=1.0\textwidth]{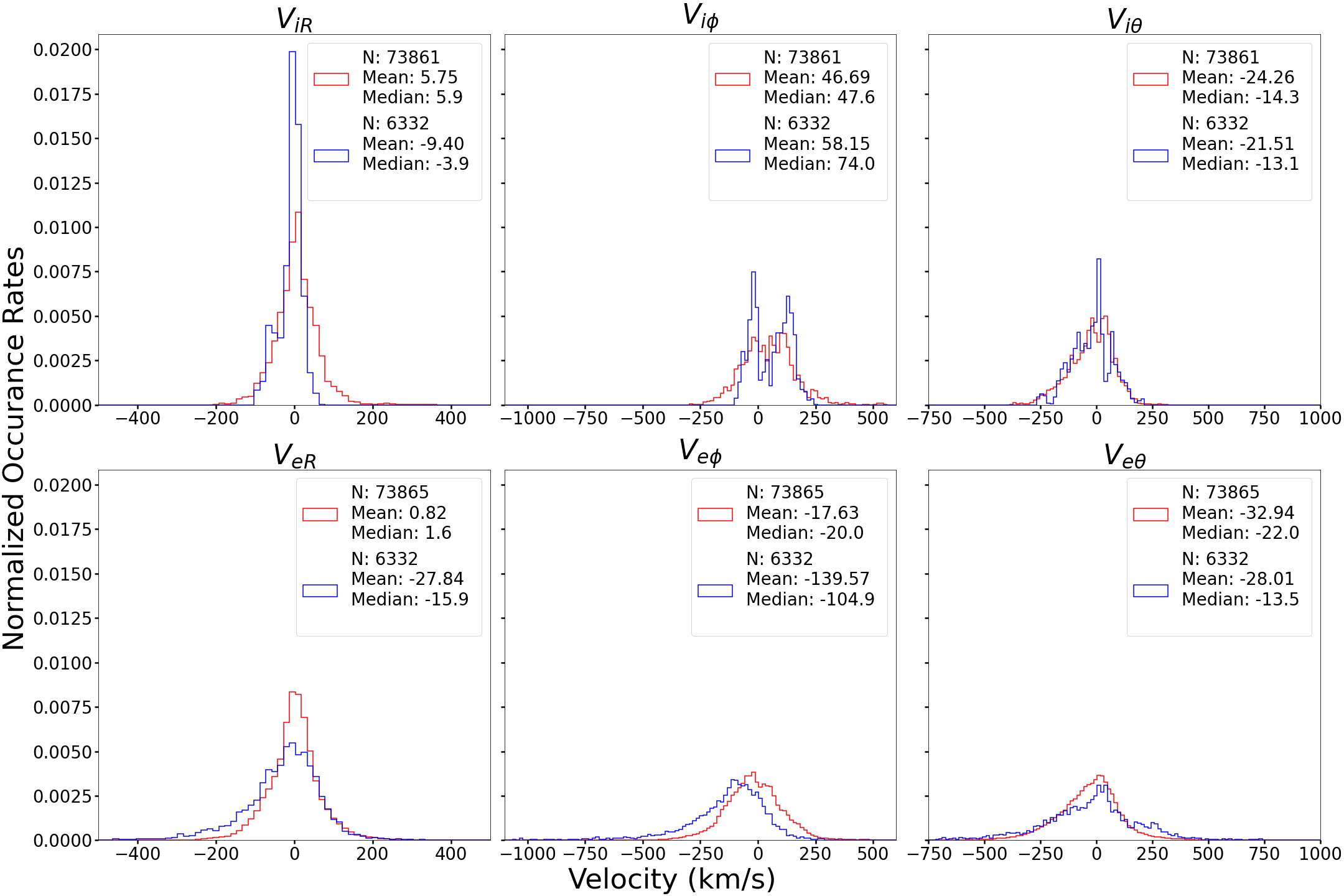}
    \caption{Ion and electron velocity histograms from the 26 EDR crossings (blue) and the 225 {regular} magnetopause crossings (red). Note the double peak structure of the EDR ion velocities in the $\hat{\phi}$ direction as well as the multi-peak structure in the $\hat{\theta}$ direction.}
    \label{fig:velocityHist}
\end{figure}

This analysis was completed for each of our currents densities ($\textbf{J}_{curl}$, $\textbf{J}_{dia \ Total_i}$, $\textbf{J}_{dia \ Total_e}$, $\textbf{J}_{dia \ \nabla N_i}$, $\textbf{J}_{dia \ \nabla \cdot T_i}$, $\textbf{J}_{dia \ \nabla N_e}$, $\textbf{J}_{dia \ \nabla \cdot T_e}$) in their component directions ($\hat{R}$, $\hat{\phi}$, $\hat{\theta}$), as shown in Table \ref{tab:table1} below. Additionally, we compiled the current data into probability distributions, which are shown in Figure \ref{fig:Jcurl_Hist} for $J_{curl}$ and its parallel and perpendicular components in the $\hat{R}$, $\hat{\phi}$, $\hat{\theta}$ directions (9 panels in total) with the EDR data points represented in the blue distributions, while the {regular} mangetopause data points are represented in the red distributions. Likewise, Figure \ref{fig:ionHist} shows the results for the ion diamagnetic current and its density and temperature components. The electron diamagnetic current and its components are provided in Figure \ref{fig:elecHist}. Figure \ref{fig:velocityHist} then shows the probability density histograms of ion and electron velocity measurements over the EDR and {regular} crossings. Each distribution figure includes labels that show the total number of points N as well as the mean and median values of their respective distributions, which are also shown in Table 1.  

\begin{table}
    \caption{Comparison of current densities obtained during the 225 {regular} and 26 EDR crossings with the following format: mean (median) $\pm$ standard error of the current densities as measured in spherical GSE coordinates ($\hat{R}$, $\hat{\phi}$, $\hat{\theta}$). The mean and median values are computed and presented in the same way as those shown in Figures \ref{fig:Jcurl_Hist} - \ref{fig:elecHist}. The EDR/{Regular} ratio was also computed based on these mean and median values.}
    \centering
    \begin{tabular}{ c c c c } 
    \hline
    Current & {Regular} $(nA/m^2)$ & EDR $(nA/m^2)$ & EDR / {Regular} \\ 
    \\
    ${J}_{R \ curl}$ & 4.96 (2.40) $\pm$ 0.39 & 30.1 (18.5) $\pm$ 2.53 & 6.1 (7.7) \\ 
    ${J}_{R \ curl_\perp}$ & -0.04 (-0.60) $\pm$ 0.33 & 14.1 (-1.70) $\pm$ 2.19 & 350 (2.8) \\ 
    ${J}_{R \ curl_\parallel}$ & 5.00 (1.00) $\pm$ 0.20 & 16.0 (6.70) $\pm$ 1.59 & 3.2 (6.7) \\
    \\
    ${J}_{R \ dia \ Total_i}$ & -1.21 (-0.70) $\pm$ 0.70 & -26.6 (-23.1) $\pm$ 3.54 & 22 (33) \\
    ${J}_{R \ dia \ \nabla N_i}$ & -7.19 (-1.20) $\pm$ 0.71 & -0.88 (-18.4) $\pm$ 2.77 & 0.12 (15) \\
    ${J}_{R \ dia \ \nabla \cdot T_i}$ & 5.98 (-0.30) $\pm$ 0.71 & -25.7 (-5.00) $\pm$ 3.15 & 4.3 (17) \\
    \\
    ${J}_{R \ dia \ Total_e}$ & -2.23 (-1.00) $\pm$ 0.10 & 13.2 (2.20) $\pm$ 0.91 & 5.9 (2.2) \\
    ${J}_{R \ dia \ \nabla N_e}$ & -2.41 (-1.00) $\pm$ 0.08 & 4.99 (-0.10) $\pm$ 0.48 & 2.1 (0.1) \\
    ${J}_{R \ dia \ \nabla \cdot T_e}$ & 0.18 (0.00) $\pm$ 0.07 & 8.25 (1.50) $\pm$ 0.74 & 49 (NA) \\
    \\ 
    ${J}_{\phi \ curl}$ & 89.5 (68.4) $\pm$ 0.52 & 349 (324) $\pm$ 3.60 & 3.9 (4.7)  \\ 
    ${J}_{\phi \ curl_\perp}$ & 56.8 (42.6) $\pm$ 0.36 & 201 (161) $\pm$ 3.06 & 3.5 (3.8) \\ 
    ${J}_{\phi \ curl_\parallel}$ & 32.7 (7.4) $\pm$ 0.39 & 148 (85.3) $\pm$ 2.69 & 4.5 (11.5) \\
    \\
    ${J}_{\phi \ dia \ Total_i}$ & 57.6 (44.4) $\pm$ 0.67 & 210 (129) $\pm$ 4.93 & 3.6 (2.9) \\
    ${J}_{\phi \ dia \ \nabla N_i}$ & 80.6 (60.8) $\pm$ 0.65 & 252 (136) $\pm$ 5.59 & 3.1 (2.2) \\
    ${J}_{\phi \ dia \ \nabla \cdot T_i}$ & -23.0 (-15.6) $\pm$ 0.63 & -42.5 (-28.7) $\pm$ 3.69 & 1.8 (1.8) \\
    \\
    ${J}_{\phi \ dia \ Total_e}$ & 6.24 (3.80) $\pm$ 0.11 & 28.1 (20.0) $\pm$ 1.08 & 4.5 (5.3) \\
    ${J}_{\phi \ dia \ \nabla N_e}$ & 5.28 (3.80) $\pm$ 0.08 & 18.3 (12.3) $\pm$ 0.60 & 3.5 (3.2) \\
    ${J}_{\phi \ dia \ \nabla \cdot T_e}$ & 0.96 (0.00) $\pm$ 0.06 & 9.82 (4.70) $\pm$ 0.86 & 10 (NA) \\
    \\
    
    ${J}_{\theta \ curl}$ & 8.90 (9.50) $\pm$ 0.57 & -38.0 (-33.1) $\pm$ 4.15 & 4.3 (4.1) \\ 
    ${J}_{\theta \ curl_\perp}$ & 1.99 (1.80) $\pm$ 0.28 & -59.4 (-45.6) $\pm$ 2.43 & 30 (25) \\ 
    ${J}_{\theta \ curl_\parallel}$ & 6.92 (5.30) $\pm$ 0.53 & 21.4 (28.0) $\pm$ 3.98 & 3.1 (5.3) \\
    \\
    ${J}_{\theta \ dia \ Total_i}$ & -6.13 (1.70) $\pm$ 0.63 & -31.1 (-16.4) $\pm$ 4.57 & 5.1 (9.6) \\
    ${J}_{\theta \ dia \ \nabla N_i}$ & -0.85 (2.60) $\pm$ 0.55 & -24.0 (-22.8) $\pm$ 4.16 & 28 (8.8) \\
    ${J}_{\theta \ dia \ \nabla \cdot T_i}$ & -5.28 (-1.70) $\pm$ 0.64 & -7.08 (17.0) $\pm$ 4.30 & 1.3 (10) \\
    \\
    ${J}_{\theta \ dia \ Total_e}$ & 1.94 (0.50) $\pm$ 0.10 & -13.2 (-6.80) $\pm$ 0.92 & 6.8 (14)\\
    ${J}_{\theta \ dia \ \nabla N_e}$ & 1.48 (0.40) $\pm$ 0.07 & -10.8 (-5.60) $\pm$ 0.48 & 7.3 (14) \\
    ${J}_{\theta \ dia \ \nabla \cdot T_e}$ & 0.46 (0.00) $\pm$ 0.06 & -2.34 (-1.80) $\pm$ 0.70 & 5.1 (NA) \\
    
    \end{tabular}
    \label{tab:table1}
\end{table}

\newpage

\subsection{EDR vs {Regular} Current Structure}

Using the {regular} events as a baseline for the background CF current sheet in the magnetopause, we can make the following observations about the magnetopause's current structure around and during EDR events from Table \ref{tab:table1} and Figures \ref{fig:Jcurl_Hist} - \ref{fig:velocityHist}: 

\begin{enumerate}
  \item EDR current densities along the primary, $\hat{\phi}$, magnetopause direction are an order of magnitude higher, on average, than a {regular} crossing. 

    \ 

    $\textbf{J}_{curl}$, $\textbf{J}_{dia \ Total_i}$, and $\textbf{J}_{dia \ Total_e}$ are all larger by an order of magnitude in the $\hat{\phi}$ direction. This applies for both the mean and median values of the measured current densities, implying that this enhancement does not just affect our data's outlying points. The amplified current density matches with general expectations of EDR crossings having strong currents because of the EDR's thin, electron-scale current sheets (see e.g. \citeA{Webster2018}). There is, however, one interesting outlier to this conclusion; $\textbf{J}_{dia \ \nabla \cdot T_i}$. Not only does the temperature-generated current density, $\textbf{J}_{dia \ \nabla \cdot T_i}$, fail to show this order of magnitude jump, but it actually exhibits the smallest increase of all the average currents, in the $\hat{\phi}$ direction, with a 1.8x increase from its {regular} event counterpart. This suggests that, while the density-generated current density, $\textbf{J}_{dia \ \nabla N_i}$, does see a substantial boost during these EDR events, $\textbf{J}_{dia \ \nabla \cdot T_i}$ does not show a similar reaction. {Note, a lack of ion heating is to be expected in the outer IDR as the Hall force cannot do work on the plasma - e.g. \citeA{Liu2022}}. Overall this indicating that ions do not see the same level of heating inside of the diffusion regions, and perhaps indicating that ions are largely unaffected by the electron-scale dynamics in the EDR.

    \ 
 
  \item There are significant {$\hat{\phi}$-directed} current signatures parallel to the local magnetic field during EDR events. 
  
    \ 
    
    $\textbf{J}_{curl}$ parallel to the local magnetic field along the primary $\hat{\phi}$ direction becomes significantly enhanced during EDR events. This can be seen in Figure \ref{fig:Jcurl_Hist} and Table \ref{tab:table1}, with the parallel component's mean value increasing by 4.5x and its median value increasing by 11.5x. While there is a significant enhancement in the $\hat{\phi}$ perpendicular current density as well, the perpendicular component's mean value increases by 3.5x and its median by only 3.8x, noticeably less than the parallel component. This difference between the parallel and perpendicular components leads to the overall amount of current parallel to the magnetic field in the primary $\hat{\phi}$ direction (along the CF current's flow) to increase in EDR events. Specifically, the mean parallel current density accounts for $42\%$ of the mean curlometer current density in EDR events, up slightly from 36\% in the {regular} events. However, looking at the more impacted median values, the median parallel current density accounts for approximately 26\% of the median curlometer current density in EDR events, over twice as much as the 11\% contribution seen in {regular} events. This indicates that a large percentage of formerly perpendicular current density - the CF current - in the {regular} magnetopause becomes parallel to the local magnetic field during EDR events. 

    \ 
    
  \item The ion diamagnetic current density dominates that of the electron current density, but to a lesser extent in EDR events. 

    \  
    
    From Table 1 and Figures \ref{fig:ionHist} and \ref{fig:elecHist}, both {regular} and EDR events show that the average $\textbf{J}_{dia \ Total_i}$ is greater than that of the average $\textbf{J}_{dia \ Total_e}$. Specifically, in {regular} crossings, the average $\textbf{J}_{dia \ Total_i}$ is 9.2x larger than $\textbf{J}_{dia \ Total_e}$. This matches with findings from \citeA{Beedle2022} where the {regular} magnetopause current was found to be ion dominated. During EDR crossings, we still find that the ions dominate, but to a lesser degree. Looking at the average values from Table 1, one can see that, during EDR crossings, $\textbf{J}_{dia \ Total_i}$ is 7.5x larger than $\textbf{J}_{dia \ Total_e}$. So, while the ions are still the main contributors, their contribution seems to decrease - primarily because of a stronger electron response in EDR events. Specifically, $\textbf{J}_{dia \ Total_e}$ sees a 4.5x average increase in EDR crossings when compared with their {regular} event counterparts, while $\textbf{J}_{dia \ Total_i}$ sees a lesser 3.6x average increase. The increasing importance of $\textbf{J}_{dia \ Total_e}$ during EDR events matches with the general expectations of an EDR crossing where the electron diffusion region itself is known to be dominated by {electron-scale} currents (e.g. \citeA{Shuster2019}). However, our results imply that, while the central electron diffusion region is dominated by these electron currents, the CF current in the magnetopause current sheet itself is still primarily ion dominated.

    \

    \item  $\textbf{J}_{dia \ Total_e}$ is composed of temperature and density components that work together instead of destructively like $\textbf{J}_{dia \ Total_i}$’s components. The enhanced $\textbf{J}_{dia \ Total_e}$ found in an EDR event comes primarily from an increase in the temperature component, whose relative contribution increases by an order of magnitude.

    \ 
    
    Figure \ref{fig:elecHist} and Table \ref{tab:table1} show that, in both EDR as well as {regular} events, the electron temperature and density components work with one other in the $+ \hat{\phi}$ direction. While this is true in both types of crossings, it is significantly more pronounced in EDR events. Additionally, the contribution of $\textbf{J}_{dia \ \nabla \cdot \overleftrightarrow{T}_e}$ is measurably enhanced in EDR crossings with a 10x increase in its average value seen in Table \ref{tab:table1} as compared to the 3.5x increase in $\textbf{J}_{dia \ \nabla {N}_e}$. This impressive enhancement to $\textbf{J}_{dia \ \nabla \cdot \overleftrightarrow{T}_e}$ is likely a result of electron heating, leading to the formation of a strong electron temperature divergence in the diffusion region. The presence of electron heating has been noted as a key component to providing pressure balance in the EDR \cite{HesseCassak2020}. Overall, the $\phi$ enhancements to both components leads to $\textbf{J}_{dia \ Total_e}$'s net strength increasing by 4.5x.
    
    \ 
    
    \item  The EDR’s ion velocity measurements are characterized by multi-peak probability distributions, while the {regular} events are described by single peak distributions. 

    \ 

    From Figure \ref{fig:velocityHist}, we can clearly see that the EDR ion velocity in the $\hat{\phi}$ and $\hat{\theta}$ directions has multi-peak distributions, which is in contrast with the single peak distributions of the {regular} events.  
    
    \ 

    Regarding the $\hat{\theta}$ ion distribution, there is a multi-peak distribution in the EDR data, which likely forms from ion outflows jets in the IDR with the jets extending out from the reconnection site along $\hat{L}$ in LMN coordinates or along $\hat{\theta}$ in our spherical coordinates. As MMS flys through the magnetopause, it can encounter both sides of the jets, forming the positive and negative peaks, or the center of the reconnection site, where there is little to no ion movement, forming the central peak near zero. {This is not the case for the {regular} events, on the other hand, as they should not pass over an active reconnection site and thus should not see both sides of the reconnection jet.}

    \

     There is also a clear dual peak in the EDR events' $V_{i \phi}$ velocity distribution. Interestingly, this double peak can be explained by considering magnetosheath flows around the dayside magnetopause. On the dusk-side of the subsolar point, the magnetosheath plasma flows in the +$\hat{\phi}$ direction, while on the dawn-side, the sheath plasma flows in the -$\hat{\phi}$ around the magnetopause. We have found that EDR events on the dusk side of the subsolar point tend to have average +$V_{i \phi}$ flows across their MP crossings - accounting for the positive peak in Figure \ref{fig:velocityHist}, while EDR events on the dawn side of the subsolar point tend to have average -$V_{i \phi}$ flows - accounting for the negative peak in Figure \ref{fig:velocityHist}. Thus, this matches with the expectation of the aforementioned magnetosheath flows. Performing a linear correlation analysis between the position of MMS along the dayside magnetopause with the average $V_{i \phi}$ across the MP current sheets gives a correlation of 0.9, which shows how strongly correlated the EDR event's location is with the appearance of these magnetosheath flows. Interestingly, this correlation is even able to be seen in the {regular} events as the linear correlation between MMS's location and average $V_{i \phi}$ is 0.78. {See Figure \ref{fig:sheathFlows} for a visual depiction of these linear correlations.} 

    \ 

    Additionally, Figure \ref{fig:velocityHist} suggests that the electron velocity in the $\hat{\phi}$ and $\hat{R}$ directions tends to be higher than the ion velocity during EDR events. This indicates periods where the current sheet is primarily controlled by electron scale current structures as was previously observed (e.g. \citeA{Phan2016}). During the {regular} crossings, the electron velocity is generally smaller than the ion velocity, indicating that the {regular} magnetopause current is dominated by the ion current, as previously reported in e.g. \citeA{Beedle2022}.

\end{enumerate}

{\section{Discussion}}

\subsection{Magnetosheath Flows in the Dayside Magnetopause}

As stated above (Item 5, Section 3.1), magnetosheath flows dominate the ion velocity running along the magnetopause boundary, or $V_{i \phi}$, in both the {regular} and EDR magnetopause current sheet as is illustrated in Figure \ref{fig:sheathFlows}. This suggests that sheath flows are primarily responsible for the ion velocity along this direction and overshadow the CF current ions in their dawn-to-dusk circulation around the dayside MP, revealing two relevant aspects of the magnetopause current system:

\begin{figure}
    \centering
    \includegraphics[width=0.9\textwidth]{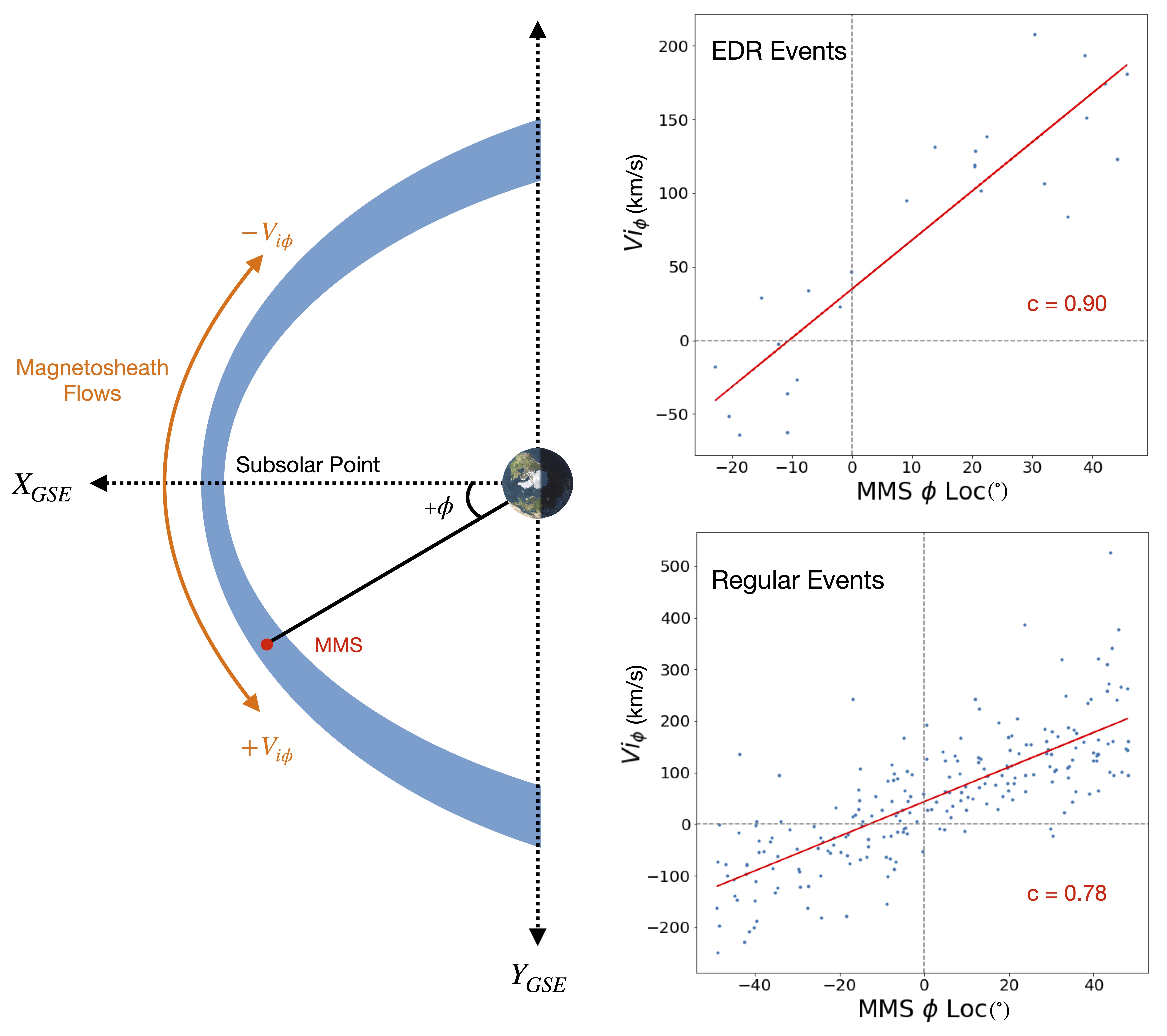}
    \caption{Left: figure illustrating the sheath flows in the + and - $\hat{\phi}$ directions around the subsolar point ($\phi$ = $0^{\circ}$) of the magnetopause. Indicated in red is an example position of MMS on its orbit around the dayside magnetopause with its location in $\phi$. Right: two linear correlation diagrams of averaged $V_{i \phi}$ over each MP crossing with MMS's $\phi$ location along the magnetopause. EDR events are represented in the top diagram while {regular} events are represented on the bottom. The correlation coefficient of the plotted linear fit (shown in red) is provided in the lower right-hand side of each plot.}
    \label{fig:sheathFlows}
\end{figure}

First, the magnetopause current sheet during both active and inactive solar wind conditions, is open to the influence of magnetosheath flows. This indicates that the current sheet, even while retaining the structure and flow mechanics of the CF current, is dominated by faster flowing sheath ions, which changes the observed $V_{i \phi}$ flow. While less correlated, the electron velocity along the magnetopause also seems to be correlated with position, with average $V_{e \phi}$ for EDR events having a correlation of 0.51 and average $V_{e \phi}$ for {regular} events having a correlation of 0.67. As both ion and electron velocities are correlated with position along the magnetopause, this means that these magnetosheath flows are likely bulk flows and should not impact the current structure of the magnetopause itself. 

Second, if we directly consider the $\hat{\theta}$ direction flows, both $V_{i \theta}$ and $V_{e \theta}$ do not show any correlation with MMS's $\phi$ position along the magnetopause. However, we can also consider MMS's location relative to the $X_{GSE}$ - $Y_{GSE}$ plane, or its $\theta$ position angle. If we consider a similar correlation analysis with the ion and electron velocities versus MMS's $\theta$ location, we see the following correlations. Both $V_{i \theta}$ and $V_{e \theta}$ for EDR events show a strong linear correlation with MMS's $\theta$ position with -0.65 and -0.6 respectively. For {regular} events, $V_{i \theta}$ and $V_{e \theta}$ show much lower correlations at -0.27 and -0.22 respectively. This shows the more open nature of the EDR event's magnetopause and also indicates the presence of sheath flows wrapping up and around the dayside magnetopause in the $\hat{\theta}$ or $Z_{GSE}$ direction.

\subsection{Current Structure in EDR Events}

As Items 1-4 of Section 3.1 suggest, EDR events depict a more complex and dynamic current structure than the {regular} magnetopause. While this is generally expected because of the added complexity from filamentary electron-scale current sheets in the EDR (e.g. \citeA{Phan2016}, \citeA{Shuster2019}, \citeA{Shuster2021}) and electron dominated Hall currents in the IDR (e.g. \citeA{Sonnerup1979}, \citeA{Nagai2001}, \citeA{Mozer2002}), there are findings that come as a surprise. The most prevalent of these is regarding the increased presence of parallel current in EDR events. Not only is this parallel current stronger than {in} the {regular} magnetopause, but also represents an interesting counterpoint to the primarily perpendicular, ion dominated diamagnetic current seen in the background magnetopause current sheet (e.g. \citeA{Beedle2022}). As the inner EDR is void of appreciable magnetic field components in the $\hat{M}$ or $\hat{\phi}$ direction (for low to no guide field cases), this $\hat{\phi}$ parallel current indicates that a measurable and significant current in this direction is detected inside the outer IDR, becoming parallel to its $\hat{M}$ directed Hall magnetic field. This could suggest additional current structure beyond the traditional 2.5D picture of the reconnection plane as is shown in zero-guide field PIC simulations such as those depicted in \citeA{Shay2016} etc. These 2.5D structures typically show strong $J_M$ generated by electron currents in the inner EDR, but whose strength diminishes inside the outer IDR where the Hall magnetic field aligns with its M direction. {This thus predicts an overall weaker parallel current structure than what our data suggests. Further investigation of this parallel current signature's mechanism, and the role that the moderate guide fields in these \citeA{Webster2018} events play, is needed however.}

\section{Summary and Conclusions}

We used MMS magnetopause crossing data over 26 dayside EDR crossings and 225 {regular} crossings to characterize differences between the diffusion regions and the background magnetopause current sheet. From this statistical analysis, we found the following: 

\begin{itemize}
    \item EDR crossings show current densities an order of magnitude higher than {regular} magnetopause crossings, representing the significantly enhanced current sheet during EDR events. 

    \ 

    \item EDR crossings contain pronounced current components parallel to the local magnetic field. This is in contrast to the primarily perpendicular current density found in the {regular} current sheet and suggests a large portion of the formerly perpendicular CF current in the {regular} mangetopause becomes parallel to the local magnetic field during EDR events.

    \ 

    \item EDR and {regular} crossings both show average ion velocities that are highly correlated with a crossing’s location along the magnetopause, indicating the presence of magnetosheath flows in the magnetopause current sheet. These flows tend to overshadow the CF current ions in their dawn-to-dusk circulation.
    
\end{itemize}

\section{Open Research}
The MMS data used in this study is publicly available from the FPI and FIELDS datasets provided at the MMS Science Data Center, Laboratory for Atmospheric and Space Physics (LASP), University of Colorado Boulder \cite{MMS_dataset}. The averaged MMS crossing data as well as the data used to create the histograms in Figures \ref{fig:Jcurl_Hist} - \ref{fig:velocityHist}, from the 225 dayside magnetopause crossings and 26 EDR events, is available through a Harvard Dataverse public database \cite{Beedle_dataset}.

\acknowledgments
The MMS current sheet database was created by Goetz Paschmann and Stein Haaland, and further developed by the International Space Science Institute Team 442, “Study of the physical processes in magnetopause and magnetosheath current sheets using a large MMS database". We thank them as well as the entire MMS team and instrument leads for the data access and support. We also thank Jim Drake, Li-Jen Chen, and Jason Shuster for our conversations regarding EDR simulations and EDR structure. We also thank the pySPEDAS team for their support and data analysis tools. This research was supported by the NASA Magnetospheric Multiscale Mission in association with NASA contract NNG04EB99C. J. M. H. B. and V. M. U. were supported through the cooperative agreement 80NSSC21M0180.


%
%




\bibliography{main.bib}

%
%
%
%
%

\end{document}